# Global phase and minimum time of quantum Fourier transform for qudits represented by quadrupole nuclei

V. P. Shauro[*] and V. E. Zobov

Kirensky Institute of Physics, Russian Academy of Sciences, Siberian Branch,
Akademgorodok 50, bld. 38, Krasnoyarsk, 660036 Russia

We demonstrate the relation between a global phase of the quantum gate and the layout of energy levels of its effective Hamiltonian required for implementing the gate for minimum time. By an example of the quantum Fourier transform gate for a qudit represented by a quadrupole nucleus with the spin $I = 1$, the effective Hamiltonians and minimum implementation times for different global phases are found. Using numerical optimal control methods, the problem of the global phase in searching for the optimal pulse shape is considered in detail for the quantum Fourier transform gate at $I = 1, 3/2, 2$, and $5/2$. It is shown that at the constrained control time the gradient algorithms can converge to the solutions corresponding to different global phases or the same global phase with different minimum times of the gate implementation.

## I. INTRODUCTION

Implementation of quantum algorithms requires performing basic quantum operations (gates) with maximum fidelity for minimum time [1, 2]. Quantum computations can be performed with the use of not only two-level (qubits) but also multilevel quantum systems (qudits) [3−5]. The latter have a number of advantages; in particular, a specified size of the computational basis is provided by fewer qudits. In implementation of quantum algorithms, one must take into account not only the operational complexity (the number of gates for algorithm execution [1]) but also the time complexity (the algorithm execution time) [6−11]. The shorter the algorithm execution time, the smaller is the loss caused by the interaction with the environment. The time complexity of quantum logic operations is determined by the quantum system and the control method used. Generally, the existence of minimum (critical) time $T_c$ for implementation of a quantum gate with an allowed error is the fundamental limitation imposed on the speed of quantum operations.

The search for effective techniques for controlling quantum systems that implement the gates with maximum fidelity for minimum time is an important problem on the way to the creation of a full-scale quantum computer. In recent years, various numerical methods have become increasingly popular that allow calculating control fields for various tasks with relatively high efficiency. The well-known numerical GRAPE [12] and Krotov [13, 14] algorithms are based on finding the minimum of a certain objective functional by using information on its gradient. Although the gradient methods yield only a local solution, these algorithms are successfully used for controlling quantum systems, since they often converge to the global minimum [15]. Moreover, in the absence of the control field constraints, the objective functional to be minimized has the only global minimum and the only global maximum, with the rest of the functional critical points being saddle points [15, 16]. However, the problem of suboptimal minima in optimization at certain control field constraints remains under-investigated [17, 18]. Here, we demonstrate that there are local minima related not only to the global phase problem but also to the existence of solutions with the same global phase but different critical times $T_c$.

---

[*] E-mail: rsa@iph.krasn.ru

As is known [7, 11], for a system with the traceless Hamiltonian, the unitary quantum gate $U_G$ can be implemented just up to the global phase factor

$$U(T) = e^{i\phi_p} U_G, \tag{1}$$

where $U(T)$ is the operator of the system evolution for time $T$. The global phase can be chosen from the set of values [7]

$$\phi_p = \phi_0 + 2\pi p / N, \quad p = 0, 1, ..., N-1, \tag{2}$$

where $N$ is the Hilbert space dimension for the system under consideration and $\phi_0$ is the smallest angle $\phi_0 \in [0, \pi]$ at which $\det\{e^{i\phi_0} U_G\} = 1$. The numerical calculations for the quantum Fourier transform (QFT) gate in the system of spins ½ showed that the minimum time for implementing the gate strongly depends on the global phase [7]. The same result was obtained for the QFT and SWAP gates on two spins ½ [11]. Finally, in study [19], the numerical simulation of the QFT on qudits with $N$ = 3 and 4 represented by quadrupole nuclei with the spins $I$ = 1 and 3/2, respectively, also showed the strong dependence of the minimum gate duration on the global phase value. In this study, we explore in more detail the effect of the global phase on the minimum time for the system with the spins $I$ = 1, 3/2, 2, and 5/2. Note that the QFT gate is often chosen for testing various control methods, since it plays a key role in many quantum algorithms [1, 2] and has rather a nontrivial form, i.e. is not straightforward to implement [7].

The paper is organized as follows. In Section II, we derive theoretical foundations for the existence of many solutions in the optimal control task for the case of a traceless Hamiltonian. By an example of the QFT gate on a qutrit ($N$ = 3) represented by a quadrupole nucleus with the spin $I$ = 1 and controlled by rf magnetic field, we find approximate analytical solutions for the control field that correspond to different global phases (2) and the same global phase but with different critical times $T_c$. In Section III, we present numerical data on the effect of the global phase on the minimum time of the QFT gate implementation on the quadrupole nucleus with the spins $I = 1, 3/2, 2$, and $5/2$ ($N = 2I +1$). Section IV contains the conclusions.

## II. GLOBAL PHASE AND EFFECTIVE HAMILTONIAN

### A. Correlation of the global phase and the effective Hamiltonian of a gate

We consider the problem of the implementation of a quantum gate in a closed quantum system with Hamiltonian

$$H(t) = H_0 + \sum_f u_f(t) H_f, \qquad (3)$$

where $H_0$ is the field-free Hamiltonian, $H_f$ is the $f$-th control Hamiltonian operator, and $u_f(t)$ is the amplitude of the corresponding control fields. We have to find the control fields $u_f(t)$ at which the operator of the system evolution for time $T$ is

$$U(T) = \hat{T} \exp\left(-i \int_0^T H(t) dt\right) \qquad (4)$$

that performs the desired logic transformation specified by the unitary matrix $U_G \in U(N)$ in a certain computational basis. Here, $\hat{T}$ is the time-ordering operator. Unitary gate $U_G$ can be presented in the exponential form

$$U_G = \exp(-iK). \qquad (5)$$

For convenience, we take the negative exponent, by analogy with the definition of evolution operator (4). Using transformation $P$, we reduce matrices $U_G$ and $K$ to the diagonal form

$$P^\dagger K P = D = \sum_{k=1}^N \lambda_k |k\rangle\langle k|,$$

$$P^\dagger U_G P = \exp(-iD) = \sum_{k=1}^N \exp(-i\lambda_k)|k\rangle\langle k|, \qquad (6)$$

where $|k\rangle\langle k|$ is the projector onto eigenstate $|k\rangle$. Now, if we add the numbers $2\pi m_k$, where $m_k$ is an integer, to one or several eigenvalues $\lambda_k$, then the value of the exponential operator in Eq. (6) does not change, but matrix $D$ changes and, consequently, matrix $K$ is transformed to the new matrix,

$$K_m \equiv K(m_1,...,m_N) = P\left(D + \sum_k 2\pi m_k |k\rangle\langle k|\right) P^\dagger = K + \sum_k 2\pi m_k P|k\rangle\langle k|P^\dagger \qquad (7)$$

with the transformed trace

$$\tilde{\Phi}_m = Tr K_m = Tr K + \sum_k 2\pi m_k. \qquad (8)$$

To implement gate $U_G$ on the quantum system with the traceless Hamiltonian [i.e., $U(T) \in \mathrm{SU}(N)$], one should take the operator

$$TH_m^{eff} = K_m - \Phi_m E, \qquad (9)$$

as an effective Hamiltonian. Here, $\Phi_m = \tilde{\Phi}_m / N$ and $E$ is the identity operator. Substituting this expression in evolution operator definition (4), we obtain

$$U_m(T) = \exp(-iTH_m^{eff}) = \exp(i\Phi_m) U_G. \qquad (10)$$

Comparing (10) and (1), we obtain $\phi_p = \Phi_m \mod(2\pi)$. Thus, different effective Hamiltonians (9) can lead to different global phases (2). Moreover, it is reasonable to suggest that there exists a set of solutions of control task (4). Different solutions corresponding to different $H_m^{eff}$ can have different critical times. Therefore, one may choose the one from the set of effective Hamiltonians $H_m^{eff}$ that has the required advantages, e.g., allows implementing gate $U_G$ in a shorter period of time.

Transformations (7) and (9) have a simple physical meaning. When in expression (7) different sets of numbers $m_k$ are chosen, effective Hamiltonian (9) changes such that one or several energy levels in it shift by $2\pi m_k / T$. The change that occurs in the average energy is eliminated by shifting the energy scale, with this average value taken for the origin of coordinates. It should be noted that transformation (7) allowed us to change the trace of matrix $K$ and, thus, pass from one global phase to another, while the unitary transformations (e.g., rotations caused by an external field) retain the matrix trace.

### B. Model system

We demonstrate the application of the above formulas for controlling a qudit represented by a quadrupole nucleus with spin $I$ in a strong static magnetic field and a control rf magnetic field. In the reference frame rotating around the static field direction (axis $z$) with rf field frequency $\omega_{rf}$ [20], the Hamiltonian acquires the form

$$H(t) = (\omega_{rf} - \omega_0) I_z + H_q + u_x(t) I_x + u_y(t) I_y, \qquad H_q = q\left(I_z^2 - \tfrac{1}{3} I(I+1)\right). \qquad (11)$$

Here, $\omega_0$ is the Larmor frequency, $I_\alpha$ is the spin projection operator along the axis $\alpha$ ($\alpha = x, y, z$), $q$ is the constant of the quadrupole interaction of a nucleus with the axially symmetric crystal field gradient, and $u_\alpha(t)$ is the projection of the control rf field onto the axis $\alpha$. We assume $\omega_{rf} = \omega_0 \gg q$. Hereinafter, the energy is measured in frequency units with $\hbar = 1$. In addition, we pass to dimensionless time and frequencies expressed in units $1/q$ and $q$, respectively.

Note that in model (11), we set the only control field with two time-dependent components along the axes *x* and *y* which simultaneously affects all the frequency transitions. In most of the simulations on multiqubit systems (see, for example, [7, 11, 17]), a set of control fields each affecting a separate qubit and not affecting the others is assumed.

In the absence of the rf field, system (11) has $N = 2I + 1$ nonequidistant energy levels for the states with different values of spin projection $I_z$:

$$|I_z = I\rangle = |1\rangle; \quad |I_z = I-1\rangle = |2\rangle; \quad \ldots \quad |I_z = -I\rangle = |N\rangle. \quad (12)$$

We choose these states as a qudit computational basis.

Below, we focus on the QFT gate implementation in the system described above. In the general case of an *N*-level system, the QFT operator in basis (12) has the form [1, 2]

$$F_N = \frac{1}{\sqrt{N}}\begin{pmatrix} 1 & 1 & 1 & \cdots & 1 \\ 1 & \sigma & \sigma^2 & \cdots & \sigma^{N-1} \\ 1 & \sigma^2 & \sigma^4 & \cdots & \sigma^{2(N-1)} \\ \vdots & \vdots & \vdots & \ddots & \vdots \\ 1 & \sigma^{N-1} & \sigma^{2(N-1)} & \cdots & \sigma^{(N-1)^2} \end{pmatrix}, \quad \sigma = \exp\left(\frac{2\pi i}{N}\right). \quad (13)$$

### C. Quantum Fourier transform on a qutrit

Before moving on to the numerical results for the QFT on system (11), we consider a simple example allowing analytical solution and helping us to understand qualitatively the findings of Section IIA. Let us consider the QFT gate for a qutrit ($N = 3$). Matrix (13) can be diagonalized by means of a sequence of selective rotations. Such sequences were explicitly found for $N = 3, 5,$ and 7 in [21] and $N = 4$ and 8 in [22]. For the gate $U_G = F_3$ (13), in expression (7) we have

$$K = -\frac{\pi}{2}\begin{pmatrix} 2g_1 & -g_2 & -g_2 \\ -g_2 & 1+g_2/2 & g_2/2 \\ -g_2 & g_2/2 & 1+g_2/2 \end{pmatrix}, \quad (14)$$

where $g_1 = \sin^2(\theta/2)$, $g_2 = \cos\theta = 1/\sqrt{3}$, and $\theta = arctg\sqrt{2}$. The diagonalizing operator is

$$P = \frac{1}{\sqrt{2}}\begin{pmatrix} \sqrt{2} & 0 & 0 \\ 0 & 1 & 1 \\ 0 & 1 & -1 \end{pmatrix}\begin{pmatrix} \sin\theta/2 & \cos\theta/2 & 0 \\ -\cos\theta/2 & \sin\theta/2 & 0 \\ 0 & 0 & 1 \end{pmatrix}. \quad (15)$$

The operators specifying the changes in the effective Hamiltonian are

$$2\pi m_1 P|1\rangle\langle 1|P^\dagger = \pi m_1\begin{pmatrix} 2g_1 & -g_2 & -g_2 \\ -g_2 & 1-g_1 & 1-g_1 \\ -g_2 & 1-g_1 & 1-g_1 \end{pmatrix},$$

$$2\pi m_2 P|2\rangle\langle 2|P^\dagger = \pi m_2 \begin{pmatrix} 2(1-g_1) & g_2 & g_2 \\ g_2 & g_1 & g_1 \\ g_2 & g_1 & g_1 \end{pmatrix}, \tag{16}$$

$$2\pi m_3 P|3\rangle\langle 3|P^\dagger = \pi m_3 \begin{pmatrix} 0 & 0 & 0 \\ 0 & 1 & -1 \\ 0 & -1 & 1 \end{pmatrix}.$$

Expressions (14) and (16) allow us to determine the effective Hamiltonian (9) required for the QFT implementation. Now, we need to find the method of its realization on system (11). The similar task for $I = 1$ was solved previously for a selective rotation operator [23]. We apply the same approach to the QFT gate. We express the effective Hamiltonian as

$$TH_m^{eff} = A + B + C, \tag{17}$$

$$A = e^{-i\varphi I_x}(H_q t_1)e^{i\varphi I_x}, \quad B = e^{-i\psi I_y}(H_q t_2)e^{i\psi I_y}, \quad C = \xi I_x + \eta I_z. \tag{18}$$

Substituting Hamiltonian (17) in the expression for the evolution operator and using the Trotter−Suzuki formula [24]

$$\left(e^{-iA/2r}e^{-iB/2r}e^{-iC/r}e^{-iB/2r}e^{-iA/2r}\right)^r = e^{-i(A+B+C)} + O(1/r^3), \tag{19}$$

we arrive at the operator product, which can be presented as the pulse sequence

$$\{\varphi\}_x \xrightarrow{t_1/2r} \{\varphi\}_{-x} \cdot \{\psi\}_y \xrightarrow{t_2/2r} \{\psi\}_{-y} \cdot \{\Omega/r\}_\Omega \cdot \{\psi\}_y \xrightarrow{t_2/2r} \{\psi\}_{-y} \cdot \{\varphi\}_x \xrightarrow{t_1/2r} \{\varphi\}_{-x}. \tag{20}$$

where $\{\theta\}_\alpha \equiv \exp(-i\theta I_\alpha)$ is the operator of nonselective rotations by angle $\theta$ around the axis $\alpha$ and $\xrightarrow{t} \equiv \exp(-it H_q)$ is the free evolution for time $t$. In the center of sequence (20), there is the rotation by the angle $\Omega = \sqrt{\xi^2 + \eta^2}$ around the axis with the direction cosines $\xi/\Omega$ and $\eta/\Omega$ along the axes $x$ and $z$, respectively. The nonselective rotations can be obtained using a simple or composite pulse of the rf field with a large amplitude [25].

Thus, to implement the QFT gate, it remains for us to determine the parameters of pulse sequence (20) by equating the sum of matrices (18) to (9). As a result, we obtain the system of equations

$$K_m - \Phi_m E = \begin{bmatrix} \tfrac{1}{6}\left((3\cos^2\varphi - 1)t_1 + (3\cos^2\psi - 1)t_2\right) + \eta & \tfrac{1}{\sqrt{2}}(-it_1\sin\varphi\cos\varphi + t_2\sin\psi\cos\psi + \xi) & -\tfrac{1}{2}(t_1\sin^2\varphi - t_2\sin^2\psi) \\ \tfrac{1}{\sqrt{2}}(it_1\sin\varphi\cos\varphi + t_2\sin\psi\cos\psi + \xi) & -\tfrac{1}{3}\left((3\cos^2\varphi - 1)t_1 + (3\cos^2\psi - 1)t_2\right) & \tfrac{1}{\sqrt{2}}(it_1\sin\varphi\cos\varphi - t_2\sin\psi\cos\psi + \xi) \\ -\tfrac{1}{2}(t_1\sin^2\varphi - t_2\sin^2\psi) & \tfrac{1}{\sqrt{2}}(-it_1\sin\varphi\cos\varphi - t_2\sin\psi\cos\psi + \xi) & \tfrac{1}{6}\left((3\cos^2\varphi - 1)t_1 + (3\cos^2\psi - 1)t_2\right) - \eta \end{bmatrix}$$

(21)

The joint solution of these equations yields the desired values of the parameters (Table I). For each value of the global phase $\phi = \{\pi/6, 5\pi/6, 9\pi/6\}$, we select the solutions with positive evolution times $t_1$ and $t_2$ that yield the minimum sum $T_m = t_1 + t_2$ and one solution with time $T_m$ next in magnitude but with the same global phase. [If we neglect the rf pulse length in (20), then the total sequence duration is $T_m = t_1 + t_2$; this value is the minimum time for implementation of the QFT gate with the use of the method under consideration. Here, we use the notation $T_m$ instead of $T_c$, because this value is only a rough estimate of critical time $T_c$ determined in Section III].

TABLE I. Parameter for implementing $H_m^{eff}$ (9) with the use of pulse sequence (20)

| $\Phi_m$ | $m_1, m_2, m_3$ | $\varphi$ | $\psi$ | $\xi$ | $\eta$ | $t_1$ | $t_2$ | $T_m$ |
|---|---|---|---|---|---|---|---|---|
| $\pi/6$ | 1, 0, 0 | $\pi/2$ | -0.905 | 0.790 | 0.105 | 3.441 | 4.267 | 7.71 |
|  | 1, 1, -1 | $\pi/2$ | -0.984 | 4.764 | 3.822 | 0.503 | 7.548 | 8.05 |
| $5\pi/6$ | 0, -1, 0 | $\pi/2$ | 0.245 | -1.431 | -1.465 | 2.409 | 0.633 | 3.04 |
|  | 0, 0, -1 | 0 | -0.963 | 2.542 | 2.251 | 2.283 | 2.688 | 4.97 |
| $9\pi/6$ | 0, 0, 0 | 0 | 1.083 | 0.320 | 0.680 | 1.077 | 2.324 | 3.40 |
|  | 0, -1, 1 | $\pi/2$ | 0.574 | -3.653 | -3.036 | 5.477 | 5.197 | 10.67 |

It can be seen from Table I that the solutions corresponding to the different global phases have different times $T_m$. Nevertheless, since we can choose arbitrary numbers $m_k$ in (7), there are many solutions that yield the same global phase (2) but different times of the gate implementation.

### III. NUMERICAL EVALUATION OF THE MINIMUM TIME FOR THE QFT
#### A. Optimization procedure

To determine the optimal control field by numerical methods, a certain iterative procedure minimizing a specific objective functional is usually used [11−14, 26]. When no constraints are imposed on the control field shape or amplitude, the error of the obtained gate is often chosen as an objective functional. The error can be determined either accurate to the global phase,

$$J_1 = \frac{1}{2} - \frac{1}{2N} \text{Re}\left\{Tr\left(U_G^\dagger U(T)\right)\right\}, \quad (22)$$

or ignoring this phase,

$$J_2 = 1 - \frac{1}{N}\left|Tr\left(U_G^\dagger U(T)\right)\right|. \quad (23)$$

Both functionals are determined in the interval [0, 1].

In this study, to find the optimal control field in (11) that minimizes gate error (22) or (23), we applied a BFGS-GRAPE algorithm [17, 26] using a standard *fminunc* function in the MATLAB package [27]. Time interval $T$ is divided into $S$ equal steps with the length $\Delta t = T/S$; the field amplitude in each $s$-th step is constant and amounts to $u_\alpha(t_s)$, where $t_s = s\Delta t$ and $s = 1, 2, \ldots, S$. For the concatenated control field vector $u = [u_x(t), u_y(t)]$, the update rule in the $k$-th algorithm iteration is given by

$$u^{(k+1)} = u^{(k)} - \beta_k \mathbf{H}_k^{-1} \nabla J_k, \quad (24)$$

where $\beta$ is the small positive parameter, $\mathbf{H}^{-1}$ is the approximate inverse Hessian defined by the BFGS formula [28], and the concatenated gradient vector $\nabla J = [\nabla J^{(x)}, \nabla J^{(y)}]$ is determined for the gate errors (22−23) as [12]

$$\nabla J_1^{(\alpha)}(t_s) = -\frac{1}{2N} \text{Re}\, Tr\left( U_G^\dagger U_S U_{S-1} \cdots \frac{\partial U_s}{\partial u_\alpha(t_s)} U_{s-1} \cdots U_1 \right) \quad (25)$$

$$\nabla J_2^{(\alpha)}(t_s) = -\frac{1}{N} \text{Re}\left\{ Tr\left( U_G^\dagger U_S U_{S-1} \cdots \frac{\partial U_s}{\partial u_\alpha(t_s)} U_{s-1} \cdots U_1 \right) Tr\left( U_1^\dagger \cdots U_S^\dagger U_G \right) \right\} \quad (26)$$

Here, the propagator $U_s = \exp[-i\Delta t H(t_s)]$ determines the system evolution in the interval $[t_{s-1}, t_s]$. The derivative of the evolution operator was calculated by the exact gradient formula in the eigenbasis of operator $U_s$ [26]:

$$\left\langle \lambda_k \left| \frac{\partial U_s}{\partial u_\alpha(t_s)} \right| \lambda_l \right\rangle = \begin{cases} -i\Delta t \left\langle \lambda_k | I_\alpha | \lambda_l \right\rangle e^{-i\Delta t \lambda_k} & \text{if } \lambda_k = \lambda_l \\ -i\Delta t \left\langle \lambda_k | I_\alpha | \lambda_l \right\rangle \dfrac{e^{-i\Delta t \lambda_k} - e^{-i\Delta t \lambda_l}}{-i\Delta t (\lambda_k - \lambda_l)} & \text{if } \lambda_k \neq \lambda_l \end{cases}$$

where $\lambda$ are the eigenvalues of $U_s$ and spin projection operator $I_\alpha$ describes the interaction with the control field in Hamiltonian (11) along the respective axis $\alpha = x, y$.

Since the gradient method converges to a local minimum, which can be different from the global one, the calculations must be repeated for many times with different guesses of initial pulse $u^{(0)}$ in (24). In our calculations for $S = 500$, the initial pulse shape was formed by setting random amplitudes in the range [−3, 3] at each 50th point of the time interval [0, $T$] with the further spline interpolation at the rest points. After that, during the optimization run, the control field amplitude is free to change without any constraints. The run is stopped when the gate error is $J^{(k)} < 10^{-8}$ or $J^{(k-1)} - J^{(k)} < 10^{-6} J^{(k-1)}$.

To determine the exact critical time for the different global phases of the gate, we calculated the dependence of the gate error on control time $T$ (Pareto front) by a Pareto front tracking (PFT) method [11]. In this technique, the optimal pulse obtained for control time $T$ is taken as an initial pulse for the optimization run for time $T \pm \Delta T$. We set $\Delta T = 0.01$. The number of time steps at passing from $T$ to $T \pm \Delta T$ was fixed; i.e., step length $\Delta t$ was varied. Such a run-to-run variation of $\Delta t$ does not significantly affect the calculation accuracy at large $S$; in our viewpoint, however, it makes the calculation more stable as compared with the regime when $\Delta t$ is fixed and $S$ is varied. The PFT technique makes it possible to significantly reduce the calculation cost and calculate the Pareto fronts for certain families of solutions with high efficiency.

### B. Numerical results

Figure 1 shows the Pareto fronts for the QFT gate on the spin $I = 1$ obtained by minimizing error $J_1$ (22). For the three possible values of the global phase, the dependences are similar, but shifted by the time scale. As critical times $T_c$, we take the values for which $J_1 < 10^{-5}$. The shortest time $T_c = 1.83/q$ corresponds to the global phase $\phi = 5\pi/6$ and the longest time $T_c = 4.44/q$ is obtained for $\phi = \pi/6$, which is qualitatively consistent with the results given in Table I. The quantitative difference is related to the fact that, considering the Hamiltonian in form (15), we significantly limited the class of possible solutions. To obtain an exact analytical solution, the explicitly time-dependent Hamiltonian should be considered.

As a rule, in practical tasks, error $J_2$ (23) independent of the global phase is minimized, since the global phase is not directly observed in the experiment. As was shown in [11], in the minimization of error $J_2$ at a fixed control time and random initial guesses, the algorithm converges to error values lying on one of the Pareto fronts (Fig. 1) corresponding to the different global phases. Suppose that such a minimization procedure yields the final operator $U(T) = e^{i\phi} U_G + \delta U$, where $\delta U$ is the matrix characterizing the error of the resulting gate. For $\|\delta U\| \ll 1$, the difference between gate errors $J_1$ (22) and $J_2$ (23) depends on $\phi$ as

$$|J_1 - J_2| \approx \sin^2(\phi/2). \tag{27}$$

Thus, if we run the minimization of gate error $J_2$ and then recalculate error $J_1$ for the final evolution operator, then we can determine the global phase of the obtained solution. We demonstrate this by the calculation of the QFT gate on the four-level system (spin $I = 3/2$). Figure 2(a) shows the diagram obtained by 500 runs for random initial pulses to minimize gate error $J_2$ (23) at $T = 2/q$. The runs are given in the order of increasing final gate error. The observed four distinct steps correspond to four different families of solutions with very similar errors. Figure 2(b) shows error $J_1$ (22) with the operator $U_G = F_4$ recalculated for the same runs. Comparing Figs. 2(a) and 2(b), we see that, in fact, the first step in Fig. 2(a) with the error $J_2 < 10^{-8}$ contains the solutions with two global phases π/8 and 9π/8, since $\sin^2(\pi/16) = 0.038$ and $\sin^2(9\pi/16) = 0.962$. The fourth step also corresponds to the solutions with the same global phases, but with larger error $J_2$. According to the findings of Section II, we suggest that the solutions that have the same global phase and different errors $J_2$ at fixed control time $T$ correspond to the solutions with the different set of numbers $m_k$ in (7) and therefore different critical times $T_c$. We will refer to the solutions with the shortest value of $T_c$ for each phase as main solutions, and the solutions with greater value of $T_c$ as secondary ones. The second step in Fig. 2(a) correspond to the main solutions with global phases 5π/8 and 13π/8 [$\sin^2(5\pi/16) = 0.691$ and $\sin^2(13\pi/16) = 0.309$], and the third step is the secondary solutions with these global phases. Thus, only about 36% of the algorithm runs ([first and second steps in Fig. 2(a)] converge to the main solutions with the various global phases. The remaining solutions are secondary ones with large critical time. It should be emphasized that the solutions with the similar error values usually have significantly different pulse shapes; thus, the formation of the steps in Figs. 2 and 4 is not related to the multiple convergence to the same pulse.

The presence of secondary solutions with the same global phase and different critical times is observed more clearly in the minimization of error $J_1$ (22) for the QFT gate with a fixed global phase, e.g. $\varphi = \pi/8$, at random initial guesses [Fig. 2(c)]. It can be seen that in this case, only about 10% of the runs converge to the main solutions with gate error $J_1 < 10^{-8}$ and two families of secondary solutions with large errors are observed at given control time. (Notice that while $J_1$ is minimizing the solutions corresponding to other global phases should have large error $J_1 \approx 1/2$ for 5π/8 and 13π/8 or $J_1 \approx 1$ for 9π/8. Therefore the probability of convergence to such solutions is extremely low, if any).

The ratio between the number of runs converging to the main and secondary solutions may differ significantly for different phases and control times $T$. It is inherent in both $J_1$ and $J_2$ scenarios of optimization. Nevertheless, for the model (11) the variation of control time $T$ did not result in a significant decrease in the probability of convergence to the secondary solutions in both scenarios.

The results of optimization shown in Fig. 2 were used in establishing the time dependence of the gate error by the PFT method for different global phases at $I = 3/2$. The Pareto fronts are shown in Fig. 3. In addition to the main solutions with the shortest critical time for each global phase, several Pareto fronts for the secondary solutions with global phases $\pi/8$ and $5\pi/8$ are shown. The shortest time $T_c = 1.62/q$ corresponds to global phases $\pi/8$ and $9\pi/8$. It can be seen in Fig. 3 that in the case $N = 4$ the Pareto fronts for global phases $\pi/8$ ($5\pi/8$) and $9\pi/8$ ($13\pi/8$) coincide pairwise within the calculation error. This also concerns the secondary solutions for the corresponding global phases. This coincidence of the critical time for the gates with the global phases different by $\pi$ reflects the general properties of half-integer spins. Indeed, since for half-integer spins the equality $e^{i(\phi+\pi)}U_G = -e^{i\phi}U_G = e^{2\pi i I_z}e^{i\phi}U_G$ is valid and the operator $e^{2\pi i I_z}$ can be realized by a strong field for a negligible time, the minimum time of the gates with different signs should be the same. The same conclusion can be made after consideration of the gates with global phases $\phi_p$ and $\phi_{p'}$ such as $|\phi_p - \phi_{p'}| = \pi$. From expression (2), we have

$$|\phi_p - \phi_{p'}| = \frac{2\pi}{N}|p - p'| = \pi \quad \Rightarrow \quad |p - p'| = \frac{N}{2}$$

Since $p$ and $p'$ are integers, this equality is valid only for even $N = 2I + 1$. For integer spins (odd $N$) it is not true and, generally, the gates with different global phases should have different critical times.

With increasing $N$ the number of solutions with different global phases and the number of secondary solutions increase. As was mentioned in [11], this can lead to the situation when the Pareto fronts for different solutions lie very close to each other. While at $N = 4$ we still can clearly distinguish the main and secondary solutions and easily grade them to the global phase (Fig. 2), already at $N = 5$ such a differentiation becomes complicated (Fig. 4). Having minimized gate error $J_1$ (22) for $N = 5$ for several hundreds of random initial pulses for each global phase and several control times $T$, we chose the runs with the minimum error and plotted the Pareto fronts for the corresponding global phases $\phi_p = \pi/5 + 2\pi p/5$, $p$=0, 1, 2, 3, 4. The result obtained is presented in Fig. 5. The shortest time $T_c = 1.90/q$ corresponds to the global phase $7\pi/5$. The figure also shows the Pareto fronts for the secondary solutions. It can be seen that after $T > 3/q$ the curves of different solutions become very close to each other. This can make certain difficulties in optimization of phase-independent gate error $J_2$ (23) for these control times, since at random initial guesses the algorithm can converge to different solutions with objective functional $J_2$ that differ by several orders of magnitude.

Figure 6 shows the results for $I = 3$ ($N = 6$). As was mentioned above, in this case the Pareto fronts for the global phases different by $\pi$ should coincide. The minor difference in such Pareto fronts (Fig. 6) is caused by numerical inaccuracy and can be eliminated by enhancing $S$ and reducing $\Delta T$. The shortest time $T_c = 2.09/q$ corresponds to the global phases $2\pi/3$ and $5\pi/3$.

Note that the secondary solutions are not observed at $I = 1$. We attribute this fact to the simplicity of a three-level system. All of the several thousands of runs for different $T$ and global phases converge very quickly to the global minimum of $J_1$ lying on a corresponding Pareto front in Fig. 1.

To verify the numerical results, some of the calculations were repeated using the Krotov algorithm with no constraints imposed on a control field [29, 30]. Despite the remarkable property of monotonic convergence, the characteristics of this algorithm are similar to those of the GRAPE first-order gradient method [30]. We observed no qualitative differences between the results obtained with the use of the Krotov and BFGS-GRAPE algorithms. Quantitative differences lie within the errors caused by discretization and features of the algorithms. In particular, the convergence rate and accuracy of the Krotov algorithm are significantly worse at the small errors $J < 10^{-4}$.

Our numerical data show that at constraints imposed on the control time, the gradient algorithms for random initial guesses can converge not only to desired solutions with a minimum gate error, but also to the secondary solutions with a much larger error. The existence of such solutions that have large critical times is consistent with the theoretical conclusions in Sec. II. Thus, the use of the gradient methods for minimizing gate error $J_1$ (22) as well as $J_2$ (23) do not guarantee the convergence to the global minimum of the functional at $U(T) \in \mathrm{SU}(N)$ and constrained control time. These conclusions may explain the traps observed in previous simulations [17, 31].

## IV. CONCLUSIONS

We showed that the same quantum gate can be implemented with the use of traceless effective Hamiltonians that differ by energy levels layout. As a consequence, the control fields implementing an effective Hamiltonian can be characterized by both the different global phases of the final gate and the minimum time for its implementation. Moreover, for a specified global phase, there can be sets of solutions with different minimum times. The methods of implementation of various effective Hamiltonians and the corresponding minimum times depend on a chosen quantum system. We chose rf-field-controlled quadrupole nuclei to implement the QFT on qudits. Minimum times for $N = 3-6$ were determined by numerical optimal control methods. For $N = 3$, additional analytical solutions for the effective Hamiltonians corresponding to different times and global phases were found. The results obtained confirm our findings.

Thus, we explained the effect of the global phase on the minimum time of the gate implementation. The obtained general relations for the change in the phase factor at the variation in the effective Hamiltonian of the gate can be useful for the systems consisting of both qubits and qudits. In construction of complex quantum circuits, the global phases discussed in this study should be controlled. Otherwise, they can spoil the interference necessary for implementation of quantum algorithms. The specific results obtained for the QFT can be used in implementation of quantum algorithms on molecules in a liquid-crystal matrix with a weak quadrupole interaction [32] and other multilevel physical systems, such as atoms controlled by laser pulses [4, 33, 34].

## ACKNOWLEDGMENTS

We thank the Referee for bring to our attention the fact that for half-integer spins the minimum times of the gates with different signs should be equal.

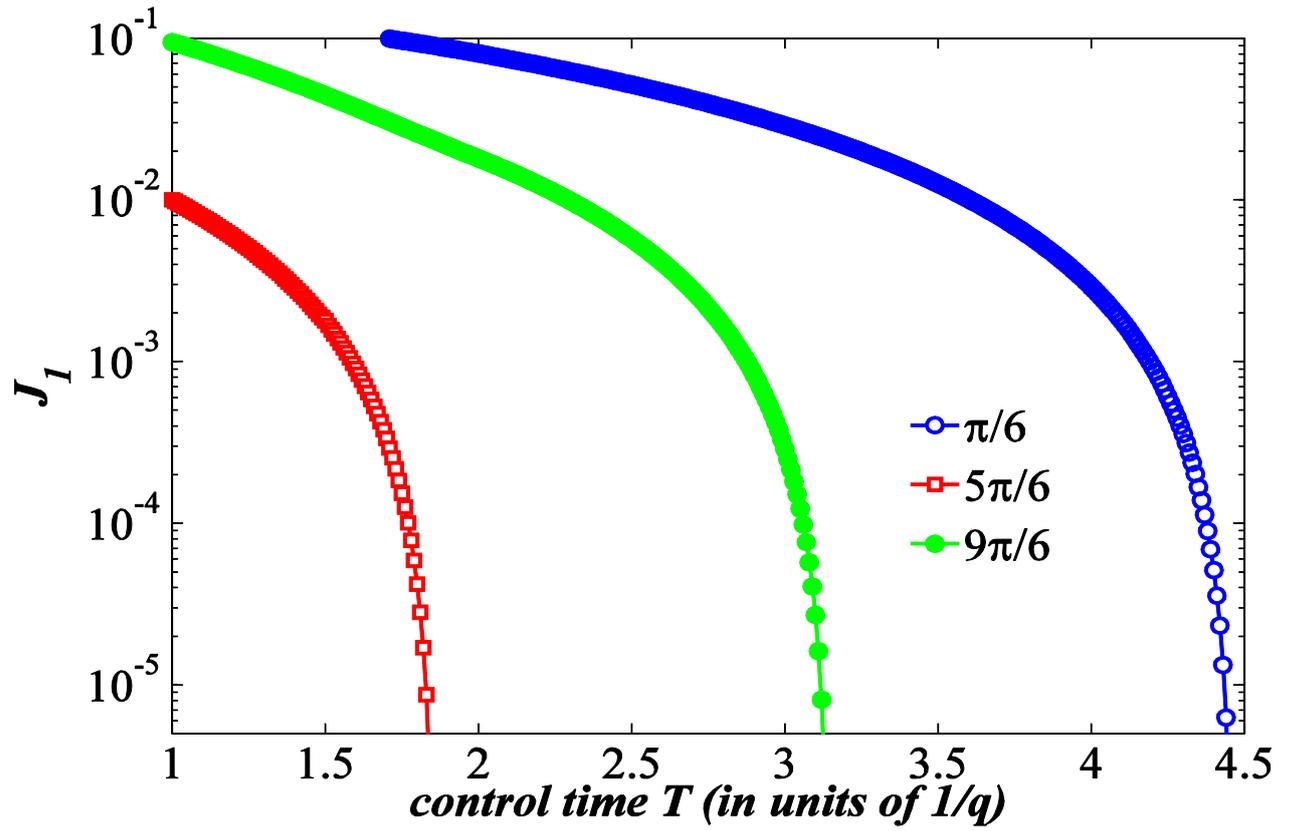

FIG. 1. (Color online) Pareto fronts for the QFT on the spin $I$=1 for three global phases of the gate.

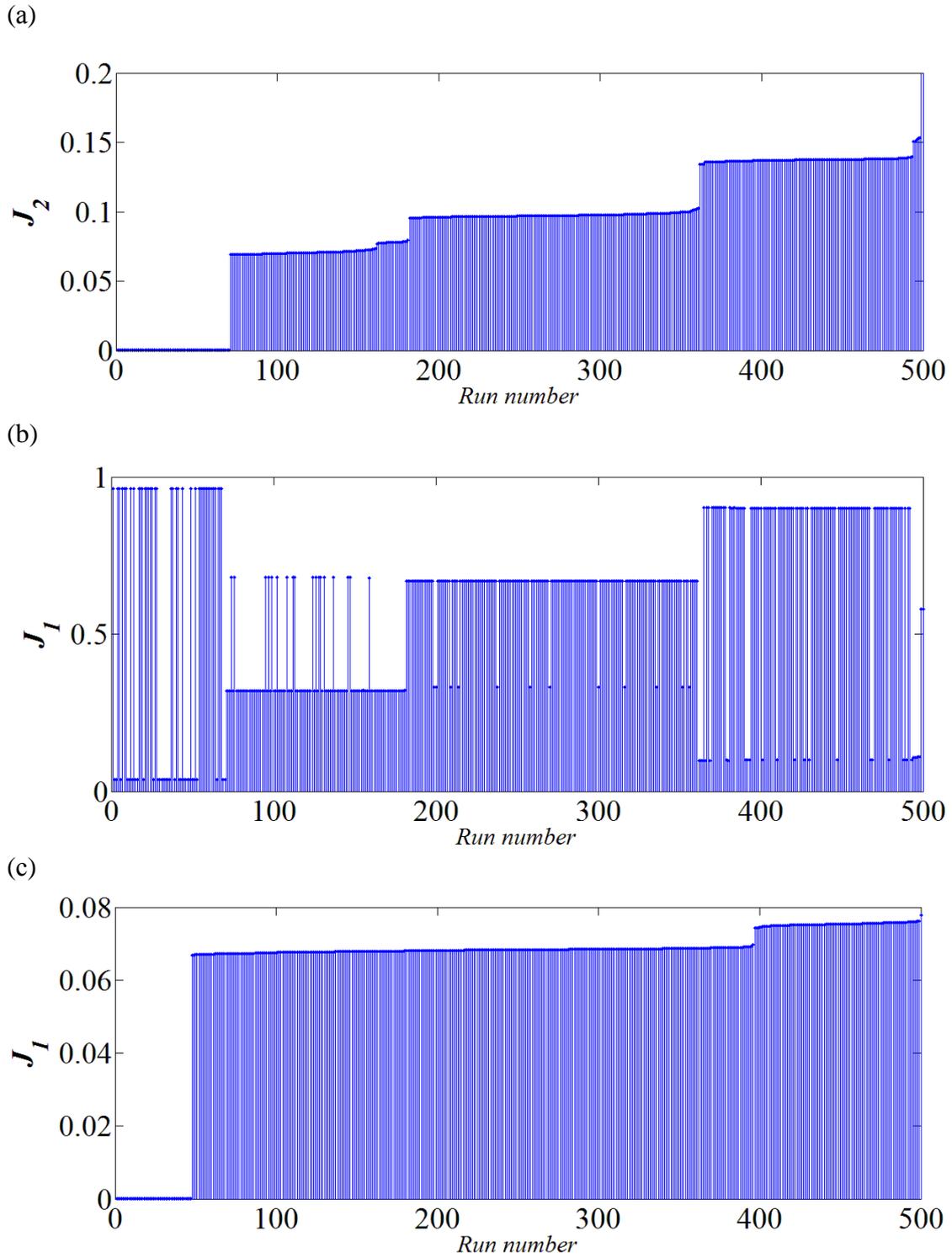

**FIG. 2.** Gate error for 500 optimization runs with random initial pulses and the control time $T = 2/q$ for the QFT on the spin $I = 3/2$. (a) Results of optimization of phase-independent gate error (23). The runs are given in the order of increasing final gate error. (b) Gate error (22) with $U_G = F_4$ for the corresponding runs in (a). (c) Results of optimization of gate error (22) with $U_G = e^{i\pi/8} F_4$. The runs are given in the order of increasing final gate error.

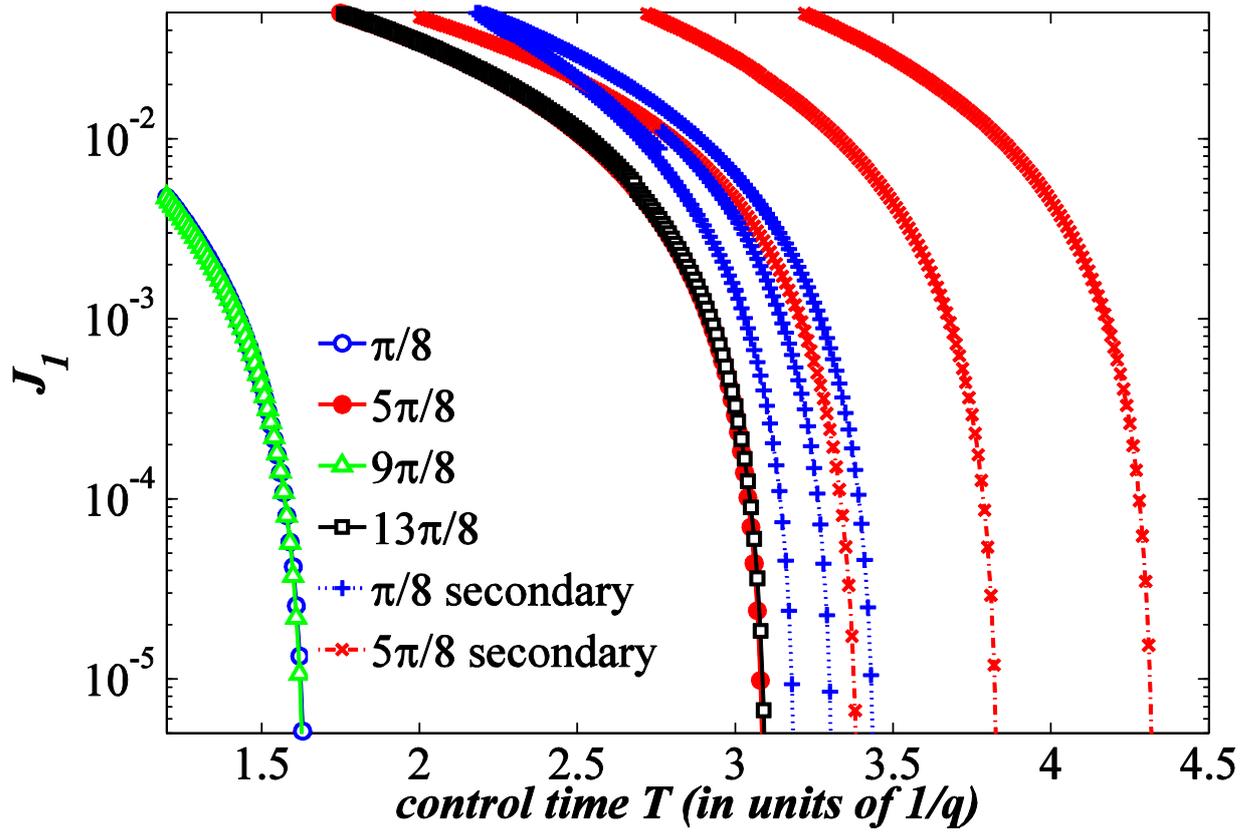

FIG. 3. (Color online) Pareto fronts for the QFT on the spin $I = 3/2$ for four global phases of the gate. Solid lines show the Pareto fronts for main solutions; dotted lines, for secondary solutions with the phase $\phi = \pi/8$; and dash-and-dot lines, for secondary solutions with the phase $\phi = 5\pi/8$.

(a)

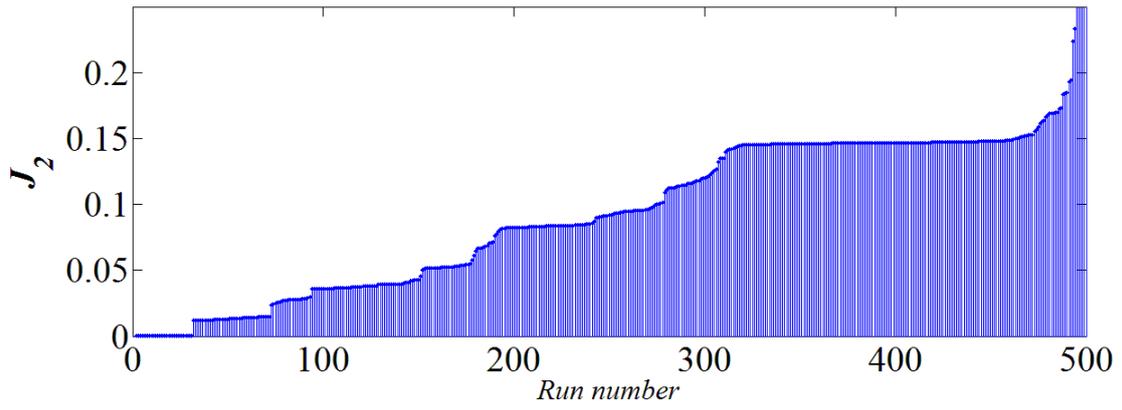

(b)

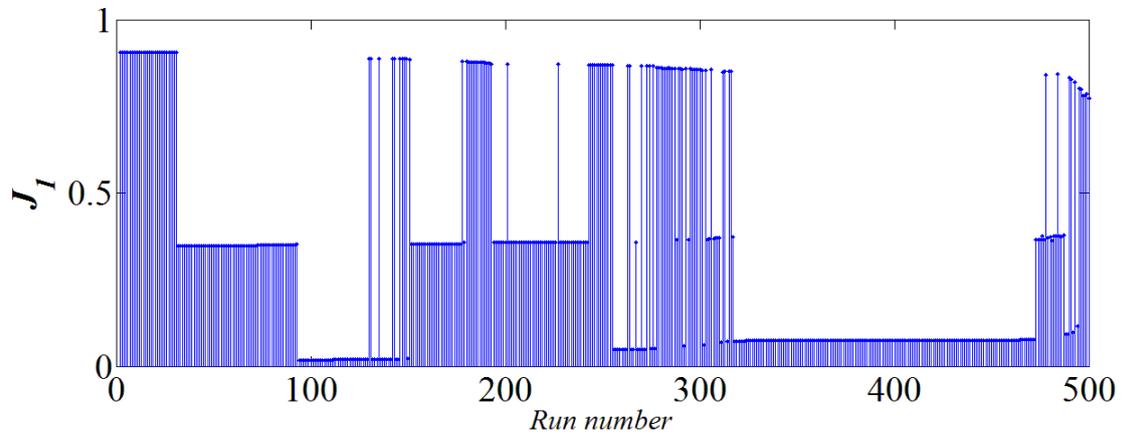

FIG. 4. Gate error for 500 optimization runs with random initial pulses and the control time $T = 2.5/q$ for the QFT on the spin $I = 2$. (a) Results of optimization of phase-independent gate error (23). The runs are given in the order of increasing final gate error. (b) Gate error (22) with $U_G = F_5$ for corresponding runs from (a).

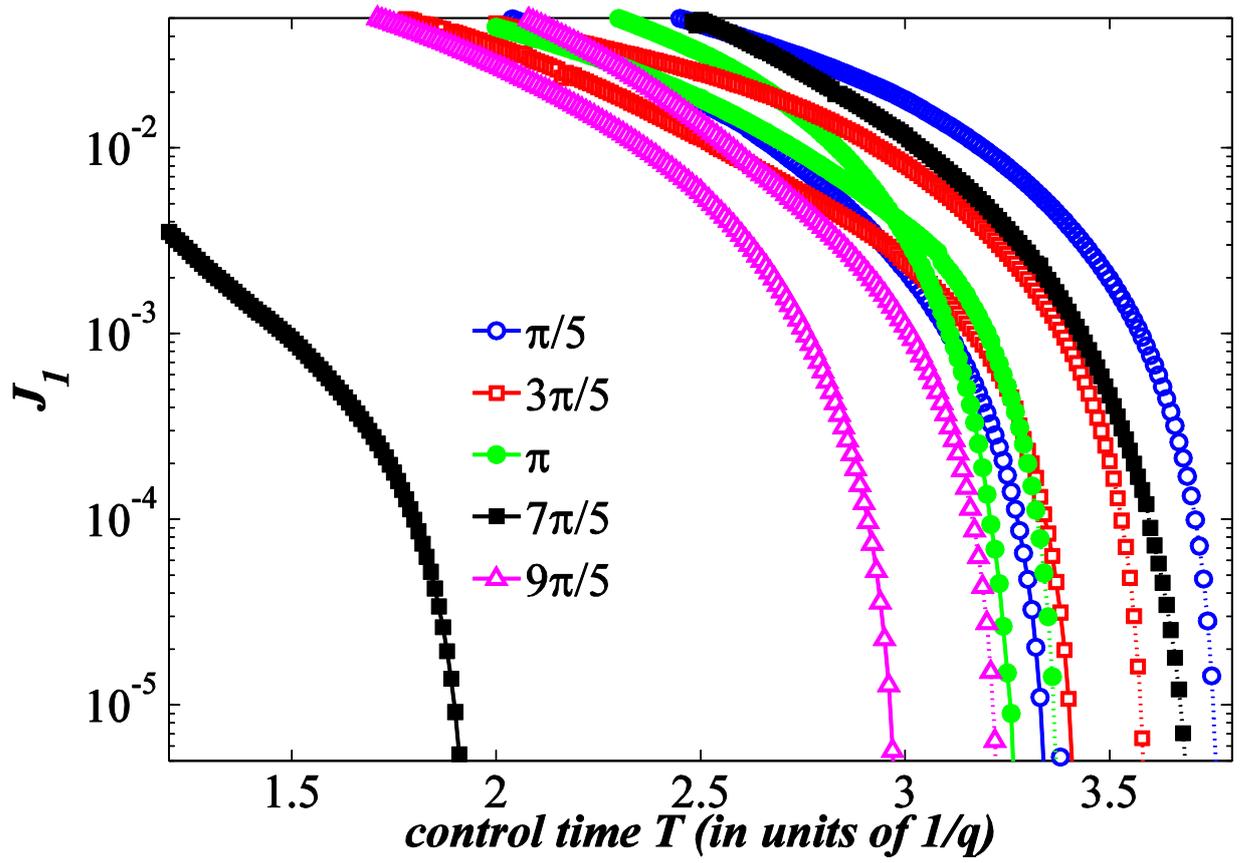

FIG. 5. (Color online) Pareto fronts for the QFT on the spin $I = 2$ for five global phases of the gate. Solid lines show the Pareto fronts for main solutions and dotted lines, for secondary solutions (one for each phase).

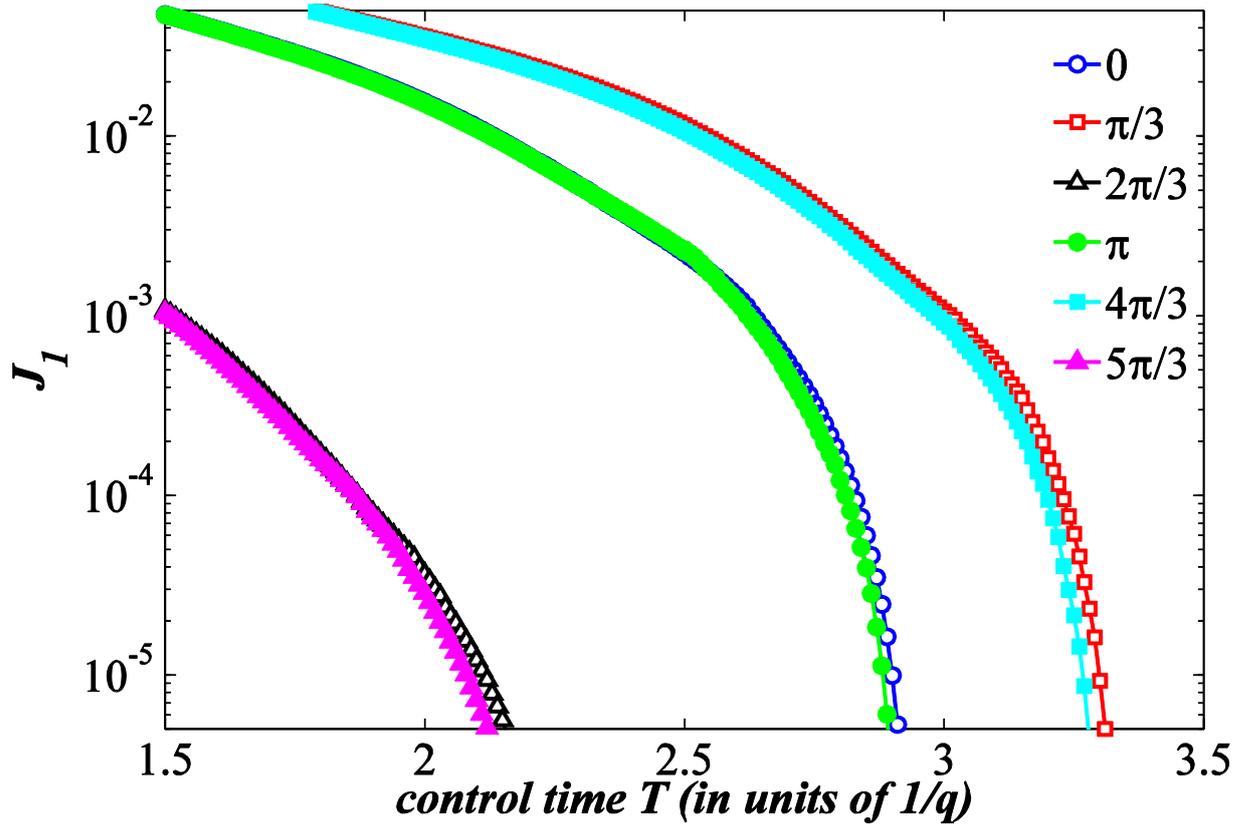

FIG. 6. (Color online) Pareto fronts for main solutions for the QFT on the spin *I* = 5/2 for six global phases of the gate.